\documentclass[aps,prd,twocolumn,superscriptaddress,showpacs,letter]{revtex4-1}            
\usepackage[utf8]{inputenc}
\usepackage{graphicx}
\usepackage{dcolumn}
\usepackage{bm}
\usepackage{cancel}
\usepackage{color}             
\usepackage{epsfig}
\usepackage{amssymb}
\usepackage{amsmath}
\usepackage{multirow}
\usepackage{hyperref}
\usepackage{cleveref}

\begin{document}

\title{Unitarity effects in elastic scattering at the LHC}
 
\author{M.~Maneyro}
\email{maneyro@fing.edu.uy}
\affiliation{Instituto de F\'{\i}sica, Facultad de Ingenier\'{\i}a, Universidad de la Rep\'ublica, \\
  J.H. y Reissig 565, 11000 Montevideo, Uruguay}
\affiliation{Department of Physics, University of Liverpool, \\ L69 7ZE, Liverpool, United Kingdom}
\author{E.~G.~S.~Luna}
\email{luna@if.ufrgs.br}
\affiliation{Instituto de F\'isica, Universidade Federal do Rio Grande do Sul, Caixa Postal 15051, 91501-970, Porto Alegre, Rio Grande do Sul, Brazil}
\author{M.~Pel\'aez}
\email{mpelaez@fing.edu.uy}
\affiliation{Instituto de F\'{\i}sica, Facultad de Ingenier\'{\i}a, Universidad de la Rep\'ublica, \\
J.H. y Reissig 565, 11000 Montevideo, Uruguay}
 

\begin{abstract}

  We study the high-energy behavior of the elastic scattering amplitude using two distinct unitarization schemes: the eikonal and the $U$-matrix. Our analysis begins with a formalism involving solely Pomerons, incorporating pion-loop insertions in the Pomeron trajectory representing the nearest singularity generated by $t$-channel unitarity. Subsequently, we explore a scenario that includes the presence of an Odderon. In our analyses, we explore the tension between the TOTEM and the ATLAS measurements for $\sigma_{tot}$ and $d\sigma/dt$ at 7, 8, and 13 TeV, and the subsequent implications for the properties of both the Pomeron and Odderon. Our results show that the Odderon phase factor $\xi_{\Bbb O}= -1$ is favored in both unitarization schemes, supporting an Odderon with a phase opposite to that of other crossing-odd components of the scattering amplitude. More interestingly, this specific phase factor stands as the sole one that aligns with results consistent with a non-zero Odderon coupling.

\end{abstract}


\maketitle

\section{Introduction}

High-energy scattering processes have been studied extensively to gain insights into the fundamental interactions between particles. Unitarization techniques play a crucial role in describing these processes.
At high energies, hadronic scattering can be described by Reggeon exchanges, as formulated by the Regge theory \cite{regge001,regge002,regge003}, in which
the behavior of the scattering amplitude is controlled by their singularities in the complex plane of angular momentum $j$.  
The simplest singularities that can occur are isolated poles at $j=\alpha(t)$, resulting in an elastic scattering amplitude written as a
series of powers of $s$,
\begin{eqnarray}
{\cal A}(s,t)=\sum_{i}\eta_{i}(t)\gamma_{i}(t)s^{\alpha_{i}(t)} ,
\label{equation01}
\end{eqnarray}
where $\eta_{i}(t)$ is the signature factor, $\gamma_{i}(t)$ is the residue function, and $\alpha_{i}(t)$ is the Regge pole trajectory, with
the index $i$ representing the $i$-th Reggeon exchange. Each term in the series represents a exchange in the $t$-channel.
The total cross section, from the optical theorem, reads
\begin{eqnarray}
\sigma_{tot}(s)=\sum_{i}4\pi g_{i} s^{\alpha_{i}(0)-1} ,
\label{equation02}
\end{eqnarray}
where $g_{i} \equiv \gamma_{i}(0)\textnormal{Im}\{ \eta_{i}(0)\}$. The Reggeon with the largest intercept is the Pomeron, namely
$\alpha_{\Bbb P}(0)=1$. However, this intercept does not allow describing the increase of all hadronic total cross sections with $s$,
theoretically predicted many years ago \cite{cheng001,cheng002,cheng003} and subsequently confirmed by experimental evidence \cite{pdg001}.
In order to describe the observed
increase of $\sigma_{tot}(s)$, the Pomeron should have a supercritical intercept given by $\alpha_{\Bbb P}(0)=1+\epsilon$ with $\epsilon > 0$.

Despite the phenomenological success of the Regge theory in describing in a unified way a large class of hadronic processes, it is worth
pointing out that the behavior of the total cross section (for $\alpha_{\Bbb P}(0)>1$) given by (\ref{equation02}) betokens the violation of the Froissart-Martin
limit at some energy scale \cite{pirner001}. In fact, contrary to the expectation that this unitarity violation would occur only far above the Large Hadron Collider (LHC) energies, an analysis of diffractive data at energies up to $\sqrt{s} = 13$ TeV has obtained $R \equiv \frac{\left| {\Bbb P\Bbb P}_{\textnormal{coupling}} \right|}{{\Bbb P}_{\textnormal{coupling}}} \simeq 0.6$ \cite{broilo001}. This value is consistent with the $R=0.5$ obtained in recent analyses conducted by Donnachie and Landshoff \cite{landshoff1a,landshoff1b}. The not-so-small values for the ratio of two-Pomeron to one-Pomeron exchange couplings clearly indicate that unitarization corrections due to double-Pomeron exchanges are already relevant at LHC energies.

The significance of the double Pomeron exchange $\Bbb P\Bbb P$ already at the LHC energy
scale anticipates the importance of processes that consider the exchange of multiple Pomerons since it is expected that unitarity can be
enforced in high-energy hadron-hadron interactions by the inclusion of the exchange series
$\Bbb P + \Bbb P\Bbb P + \Bbb P \Bbb P\Bbb P + \ldots $. In this context, the Pomeron intercept is an effective power, representing not only the exchange of a single Pomeron, but also $n$-Pomeron exchange processes, $n\geq 2$. 
As a consequence, we have a slight decrease of $\epsilon$ with $s$ since multiple exchanges must tame the rise of $\sigma$.
However, despite the advances in understanding the nature of the Pomeron in the last decades, we still need to learn how to fully compute the contributions to the scattering amplitude from multiple-Pomeron exchange processes with $n \geq 3$.
On the other hand, it is well-established that some unitarization schemes sum appropriately rescattering diagrams representing the exchange of several particular multiparticle states. These schemes are primarily based on phenomenological arguments. They are effective procedures for taking into account many of the properties of unitarity in the $s$-channel or, at the very least, for preventing the Froissart-Martin bound for $\sigma_{tot}$ from being violated.

This paper focuses on two unitarization schemes: the eikonal and the $U$-matrix. We are interested in studying the unitarization effects on the behavior of proton-proton ($pp$) and antiproton-proton ($\bar{p}p$) total cross sections at LHC and cosmic rays energies. Since we are dealing with energy scales close to the unitarity bound, it is imperative to define the domains of validity of the various unitarization schemes.
Moreover, the LHC has performed precise measurements of hadronic processes that provide a unique constraint on the behavior of the
scattering amplitude and on the nature of unitarization at high energies.

In anticipation of the forthcoming discussion in the next section, it is important to acknowledge that the unitarization procedure depends on choosing input amplitudes in the Born approximation that satisfy certain physical principles. In our case, we utilize Regge theory as the basis for the input amplitudes. Along with the typical soft Pomeron, we also investigate the presence of an Odderon since the results of the TOTEM Collaboration at the LHC reveal the manifestation of a $C=-1$ counterpart to the $C=+1$ Pomeron.

The outline of this paper is as follows. In Sec. II, we introduce the two standard unitarization schemes to be used, where we deduce the limits imposed by each scheme and their unitarity-equation solutions in the impact parameter representation. Sec. III presents the Regge formalism used to model the input Born-level amplitudes. This section introduces a nonlinear term in the Pomeron trajectory coming from the nearest $t$-channel singularity. In Sec. IV, we introduce the Odderon contribution to the Born amplitude, investigating the role of its phase factor in the two unitarization schemes. Finally, in Sec. V, we draw our results and conclusions.

\section{Unitarization Schemes}

In practice, hadronic diffraction can be described by analogy with the theory of diffraction in classical optics. From Babinet's
principle, the diffraction pattern from an absorptive disc is identical to that from a circular aperture; in the case of high-energy
particle scattering, the optical analog is the diffraction by an opaque obstacle. The short wavelength condition
$kR \gg 1$, where $k$ is the wave number and $R$ is the dimension of the hadron radius, as well the large distance condition $R/D \ll 1$,
where $D$ is the distance at which the scattered hadrons are observed behind the scatterer, are the typical conditions found in high
energy hadron scattering. Furthermore, the condition that characterizes the Fraunhofer diffraction regime, $kR^{2}/D \ll 1$, is also
satisfied. In this regime, the elastic scattering amplitude reads
\begin{eqnarray}
{\cal A}(s,t) = s \int_{0}^{\infty}b\, db\, J_{0}(b\sqrt{-t})\, H(s,b) ,
\label{scatter002}
\end{eqnarray}
where $H(s,b)$ is the elastic profile function, and $-t = q^{2}$. The equation (\ref{scatter002}) is nothing more but the partial wave
expansion for a spherically symmetric
potential in the limit of high energy ($k\to \infty$) and small scattering angles ($\varphi \to 0$), where large values of the angular
momentum $l$ (namely $l \to \infty $) are important;
the sum over many different partial-waves with large $l$ leads us to rewrite the integral over $l$ as an integral representation in the
$b$-space, where $b$ is the impact parameter \cite{regge001,regge003}.
In this picture, the physical interpretation of the elastic profile function is straightforward: $H(s,b)$ describes how much absorption
the hadron incident wave has suffered, caused by the opening of inelastic channels at a given $b$.

Certain distinctive features of the high-energy scattering amplitude, including phenomena like shadowing and saturation, among others, are better illuminated when examined in the impact parameter $b$-representation \cite{giffon001}. In $b$-space the $s$-channel unitarity equation reads \cite{regge001}
\begin{eqnarray}
2 \, \textnormal{Im}\, H(s,b) = |H(s,b)|^{2} + G_{in}(s,b) ,
\label{unitarity001}
\end{eqnarray}
where $G_{in}(s,b)$, a real non-negative quantity, is the inelastic overlap function.
Integrated over two-dimensional impact parameter space, equation (\ref{unitarity001}) reduces to the ordinary relation between the
total, elastic, and inelastic cross sections, 
\begin{eqnarray}
\sigma_{tot}(s) = \sigma_{el}(s) + \sigma_{in}(s),
\label{unitarity011}
\end{eqnarray}
which is normalized such that for the total, elastic and inelastic cross section holds
\begin{eqnarray}
  \sigma_{tot}(s) = \frac{4\pi}{s} \textnormal{Im}\,{\cal A}(s,t=0)= 2\pi \int_{0}^{\infty} b \, db \, 2 \, \textnormal{Im} H(s,b) ,
  \nonumber \\
\label{integrated001}
\end{eqnarray}
\begin{eqnarray}
  \sigma_{el}(s) = \frac{\pi}{s^{2}} \int_{-\infty}^{0} dt \, \left| {\cal A}(s,t)  \right|^{2} = 2\pi \int_{0}^{\infty} b \, db
  \, \left| H(s,b) \right|^{2} , \nonumber \\
\label{integrated002}
\end{eqnarray}
\begin{eqnarray}
  \sigma_{in}(s) = 2\pi \!\! \int_{0}^{\infty} \!\! b \, db
  \, \left( 2 \, \textnormal{Im} H(s,b) - \left| H(s,b) \right|^{2} \right) . 
\label{integrated003}
\end{eqnarray}

After defining a function $\rho (s,b)$ as the ratio of real to imaginary parts of $H(s,b)$,
\begin{eqnarray}
\rho (s,b) = \frac{\textnormal{Re}\, H(s,b)}{\textnormal{Im}\, H(s,b)} ,
\label{unitarity002}
\end{eqnarray}
and solving the quadratic equation for $\textnormal{Im}\, H(s,b)$ resulting from (\ref{unitarity001}), we get
\begin{eqnarray}
\textnormal{Im}\, H(s,b) = \frac{1 \pm \sqrt{1 - \left( 1 + \rho^{2}  \right) G_{in}(s,b)}}{1 + \rho^{2}} ;
\label{unitarity003}
\end{eqnarray}
from (\ref{unitarity003}), we see that $G_{in}(s,b)$ must fit onto the interval
\begin{eqnarray}
0 \leq G_{in}(s,b) \leq (1 + \rho^{2})^{-1} ,
\label{unitarity004}
\end{eqnarray}
where we have required $\textnormal{Im}\, H(s,b)$ be real.

The construction of unitarized scattering amplitudes relies on two formal steps. First, the choice of a Born term (or input term),
namely a Born scattering amplitude in momentum space, ${\cal F}(s,t)$, with the crossing-even and crossing-odd parts defined as
\begin{eqnarray}
{\cal F}^{\pm}(s,t) = \frac{1}{2} \left[ {\cal F}^{pp}(s,t) \pm  {\cal F}^{\bar{p}p}(s,t) \right] .
\label{eqn01} 
\end{eqnarray}
The correspondent crossing-even and crossing-odd Born amplitudes in $b$-space are given by
\begin{eqnarray}
h^{\pm}(s,b) = \frac{1}{2s}\int_{0}^{\infty} q \, dq \, J_{0}(bq) {\cal F}^{\pm}(s,-q^{2}) . 
\label{eqn02}
\end{eqnarray}
For practical convenience, it is valuable to consider the alternative input representation $\chi^{\pm} (s,b)$, which is defined as $\chi^{\pm} (s,b) \equiv 2h^{\pm}(s,b)$.

The second step consists of writing the scattering amplitude $H^{pp}_{\bar{p}p}(s,b)$ in terms of the Born amplitudes
$\chi^{pp}_{\bar{p}p}(s,b)$.
Once this is done, the amplitude in momentum space, to be used in calculating the observables, is finally obtained from
the inverse Fourier-Bessel transform of $H^{pp}_{\bar{p}p}(s,b)$:
\begin{eqnarray}
{\cal A}^{pp}_{\bar{p}p}(s,t) = s \int_{0}^{\infty}b\, db\, J_{0}(bq)\, H^{pp}_{\bar{p}p}(s,b) .
\label{scatter007}
\end{eqnarray}
Notice that we recover the elastic scattering amplitude in Fraunhofer's approximation, equation (\ref{scatter002}), as expected.
The precise form of the functional $H[\chi(s,b)]$ depends on the choice of the unitarization scheme to be applied.
The schemes are related to the appropriate solutions of the unitarity equation (\ref{unitarity003}).

The eikonal unitarization corresponds to the solution of equation (\ref{unitarity003}) with the minus sign, i.e., the one with the
negative square root. The eikonal scheme (Es) leads us to the relation
\begin{eqnarray}
H(s,b) = i \left[ 1 - e^{i\chi (s,b)}  \right] ,
\label{eikrelation01}  
\end{eqnarray}
so that
\begin{eqnarray}
{\cal A}_{[Es]}(s,t) = is \int^{\infty}_{0} b\, db\, J_{0}(bq) \left[ 1 - e^{i\chi (s,b)} \right] .
\end{eqnarray}
From equations (\ref{unitarity003}) (with the negative square root) and (\ref{unitarity004}),  we obtain an upper limit on the imaginary
part of $H(s,b)$,
\begin{eqnarray}
0 \leq \textnormal{Im} H(s,b) \leq (1 + \rho^{2})^{-1} ,
\end{eqnarray}
while solving the unitarity equation (\ref{unitarity001}) for $G_{in}(s,b)$ in terms of the eikonal function
(see equation (\ref{eikrelation01})) yields
\begin{eqnarray}
G_{in}(s,b) = 1 - e^{-2\, \textnormal{Im}\,\chi(s,b)} ;
\end{eqnarray}
the positivity condition on $G_{in}(s,b)$ and the upper limit on $\textnormal{Im} H(s,b)$ restrict the imaginary part of $\chi(s,b)$ over
the range
\begin{eqnarray}
0 \leq \textnormal{Im} \chi(s,b) \leq -\frac{1}{2}\ln \left( \frac{\rho^{2}}{1+\rho^{2}} \right) .
\end{eqnarray}

In the limit of a perfectly absorbing profile, namely $\rho =0$, the functions $H(s,b)$ and $\chi (s,b)$ are purely imaginary (and the scattering amplitude ${\cal A}(s,t)$ also becomes imaginary), with the asymptotic result that
$\textnormal{Im} H(s,b) = \left| H(s,b) \right|^{2} = 1$. When inserted into equations (\ref{integrated001}) and (\ref{integrated002}),
these results lead us to the asymptotic behavior $\sigma_{el}/\sigma_{tot}=1/2$.

The $U$-matrix unitarization corresponds to the solution of the unitarity equation (\ref{unitarity003}) with the plus sign, i.e., the one
with the positive square root. The $U$-matrix scheme (Us) leads us to the relation
\begin{eqnarray}
H(s,b) = \frac{\chi(s,b)}{1-i\chi(s,b)/2} ,
\label{unitarz01}
\end{eqnarray}
so that
\begin{eqnarray}
{\cal A}_{[Us]}(s,t) = i s \int^{\infty}_{0} b\, db\, J_{0}(bq) \left[ \frac{2 \chi(s,b)}{\chi(s,b) + 2i} \right] .
\end{eqnarray}
From equations (\ref{unitarity003}) (with the positive square root) and (\ref{unitarity004}),  we see that the imaginary
part of $H(s,b)$ is constrained to lie in the interval
\begin{eqnarray}
(1 + \rho^{2})^{-1} \leq \textnormal{Im} H(s,b) \leq 2 (1 + \rho^{2})^{-1} ,
\end{eqnarray}
while in the black disc and $\rho \to 0$ limits, we have $\textnormal{Im} H(s,b) = 2$ and $\left| H(s,b) \right|^{2} = 4$. When inserted into
equations (\ref{integrated001}) and (\ref{integrated002}), these results lead us to the asymptotic behavior $\sigma_{el}/\sigma_{tot}=1$.
Thus, $H(s,b)$ may exceed the black disc limit in this approach. 

It is worth noting that both unitarization schemes have a common feature of mapping the input amplitude $\chi(s,b)$ onto the unitarity
circle \cite{cudell01,cudell02}.
Equally important is the fact that both of these unitarization schemes are expected to converge to the same scattering amplitude at low
energies \cite{selyugin01}. However, as we delve into higher energy regimes, each scheme incorporates distinct higher-order diagrams,
signifying different underlying scattering mechanisms. Consequently, this divergence in the unitarization approach implies potential
disparities between the eikonal and $U$-matrix unitarized amplitudes at high energies, leading to observable variations in quantities such
as the cross section.

\section{Born amplitudes: Pomeron input}

The input amplitudes ${\cal F}_{i}(s,t)$ are derived from the Regge formalism, where the scattering amplitudes are associated with
Reggeon exchange amplitudes. The corresponding amplitudes in the impact parameter space, $\chi_{i}(s,b)$, are obtained through a
Fourier-Bessel transform applied to ${\cal F}_{i}(s,t)$. Specifically, equations (\ref{eqn01}) and (\ref{eqn02}), along with the
relationship $\chi^{\pm} (s,b) \equiv 2h^{\pm}(s,b)$, allow us to express $\chi_{i}(s,b)$ as follows:
\begin{eqnarray}
\chi_{i}(s,b) = \frac{1}{s}\int \frac{d^{2}q}{2\pi}\, e^{i{\bf q}\cdot {\bf b}} \, {\cal F}_{i}(s,t) .
\label{gtgt01}
\end{eqnarray}
In this paper we explore different combinations of Reggeons and assess their effectiveness in describing high-energy data.
We begin by examining the input amplitudes ${\cal F}_{-}(s,t)$, ${\cal F}_{+}(s,t)$, and ${\cal F}_{\Bbb P}(s,t)$, where
${\cal F}_{-}(s,t)$ represents the exchange of Reggeons having $C=-1$ parity ($\omega$ and $\rho$), ${\cal F}_{+}(s,t)$ the exchange of
Reggeons having $C=+1$ parity ($a_{2}$ and $f_{2}$), while ${\cal F}_{\Bbb P}(s,t)$ represents the $C=+1$ Pomeron exchange.

The physical amplitudes in the impact parameter space are obtained by summing the Fourier-Bessel transforms of all possible exchanges. This summation is represented as
\begin{eqnarray}
\chi^{pp}_{\bar{p}p} (s,b) =  \chi_{\Bbb P}(s,b)  + \chi_{+}(s,b)  \pm \chi_{-}(s,b) .
\label{eikonal008}
\end{eqnarray}
Specifically, the Born amplitude for each single exchange is
\begin{eqnarray}
{\cal F}_{i}(s,t) = \beta_{i}^{2}(t)\eta_{i}(t)\left( \frac{s}{s_{0}} \right)^{\alpha_{i}(t)} ,
\label{equation05}
\end{eqnarray}
$i=-, +, \Bbb P$, where $\eta_{i}(t)$ is the signature factor, $\beta_{i}(t)$ is the elastic proton-Reggeon vertex, $\alpha_{i}(t)$ is the
Regge trajectory, and  $s_{0} \equiv 1$ GeV$^{2}$ is an energy scale. For trajectories with odd-signature
$\eta_{i}(t)=-ie^{-i\frac{\pi}{2}\alpha_{i}(t)}$, whereas for those with even-signature $\eta_{i}(t)=-e^{-i\frac{\pi}{2}\alpha_{i}(t)}$.
The Reggeons with positive charge-conjugation are taken to have an exponential form for the proton-Reggeon vertex,
\begin{eqnarray}
\beta_{+}(t)=\beta_{+}(0)\exp (r_{+}t/2 ) ,
\end{eqnarray}
and to lie on a linear trajectory of the form
\begin{eqnarray}
\alpha_{+}(t) = 1 - \eta_{+} + \alpha^{\prime}_{+} t.
\end{eqnarray}
Similarly, the Reggeons with negative charge-conjugation are described by the pa\-ra\-me\-ters
$\beta_{-}(0)$, $r_{-}$, $\eta_{-}$, and $\alpha^{\prime}_{-}$.

For Pomeron exchange, we adopt the nonlinear trajectory
\cite{anselm001,kmr001,kmr002,kmr003},
\begin{eqnarray}
\alpha_{\Bbb P}(t) = \alpha_{\Bbb P}(0) + \alpha^{\prime}_{\Bbb P} t + \frac{m_{\pi}^{2}}{32\pi^{3}}\,
h(\tau) ,
\label{pomnlin}
\end{eqnarray}
where $\alpha_{\Bbb P}(0) = 1 + \epsilon$ and
\begin{eqnarray}
h (\tau) &=& -\frac{4}{\tau}\, F_{\pi}^{2}(t) \left[  2\tau - (1+\tau)^{3/2} \ln \left( \frac{\sqrt{1+\tau}+1}{\sqrt{1+\tau}-1} \right)
\right. \nonumber \\
& & + \left. \ln \left( \frac{m^{2}}{m_{\pi}^{2}} \right) \right]  ,
\label{nonlinear01}
\end{eqnarray}
with $\epsilon > 0$, $\tau = 4m_{\pi}^{2}/|t|$, $m=1$ GeV, and $m_{\pi}=139.6$ MeV. In equation (\ref{nonlinear01}), $F_{\pi}(t)$ is the form
factor of the pion-Pomeron vertex, for which we take the usual pole expression $F_{\pi}(t)=\beta_{\pi}/(1-t/a_{1})$.
Here the coefficient $\beta_{\pi}$ specifies the value of the pion-Pomeron coupling and for this we take the additive quark model relation
$\beta_{\pi}/\beta_{I\!\!P}(0)=2/3$.
The third term on the right-hand side of (\ref{pomnlin}) corresponds to pion-loop insertions and is generated by $t$-channel unitarity.

We investigated two different forms for the proton-Pomeron vertex. This investigation became particularly pertinent with the emergence of the LHC, as it allowed us to assess the efficacy of these vertex models in accurately describing high-energy experimental data. The first vertex, specifying our ``Model I'', is given by the exponential
form
\begin{eqnarray}
\beta_{\Bbb P}(t)=\beta_{\Bbb P}(0)\exp \left( \frac{r_{\Bbb P}t}{2} \right) .
\label{vertex01}
\end{eqnarray}

The second proton-Pomeron vertex, referred to as ``Model II'', has the power-like form
\begin{eqnarray}
\beta_{\Bbb P}(t)=\frac{\beta_{\Bbb P}(0)}{(1-t/a_{1})(1-t/a_{\Bbb P})} ;
\label{vertex02}
\end{eqnarray}
note that the free parameter $a_{1}$ in (\ref{vertex02}) is the same as the one in the expression for $F_{\pi}(t)$.
In the forthcoming analyses, we fix this parameter at $a_{1}=m_{\rho}^{2} = (0.776\, \textnormal{GeV})^{2}$ \cite{kmr004}.

The total cross section, the elastic differential cross section, and the $\rho$ parameter are expressed in terms of the physical
amplitude ${\cal A}^{pp}_{\bar{p}p}(s,t)$,
\begin{eqnarray}
\sigma^{pp, \bar{p}p}_{tot}(s)=\frac{4\pi}{s}\, \textnormal{Im}\, {\cal A}^{pp}_{\bar{p}p}(s,t=0) ,
\label{ertion001}
\end{eqnarray}
\begin{eqnarray}
\frac{d\sigma^{pp, \bar{p}p}}{dt}(s,t)=\frac{\pi}{s^{2}}\, \left| {\cal A}^{pp}_{\bar{p}p}(s,t) \right|^{2} ,
\label{ertion002}
\end{eqnarray}
\begin{eqnarray}
\rho^{pp, \bar{p}p}(s)=\frac{\textnormal{Re}\, {\cal A}^{pp}_{\bar{p}p}(s,t=0)}{\textnormal{Im}\, {\cal A}^{pp}_{\bar{p}p}(s,t=0)} ,
\label{ertion003}
\end{eqnarray}
together with the replacements  ${\cal A}^{pp}_{\bar{p}p}(s,t) = {\cal A}^{pp, \bar{p}p}_{[Es]}(s,t)$ or ${\cal A}^{pp, \bar{p}p}_{[Us]}(s,t)$,
where
\begin{eqnarray}
{\cal A}^{pp, \bar{p}p}_{[Es]}(s,t) = is \int^{\infty}_{0} b\, db\, J_{0}(bq) \left[ 1 - e^{i\chi^{pp}_{\bar{p}p} (s,b)} \right] 
\end{eqnarray}
or
\begin{eqnarray}
{\cal A}^{pp, \bar{p}p}_{[Us]}(s,t) = i s \int^{\infty}_{0} b\, db\, J_{0}(bq) \left[ \frac{2 \chi^{pp}_{\bar{p}p}(s,b)}{\chi^{pp}_{\bar{p}p}(s,b) + 2i} \right] .
\end{eqnarray}

\section{Born amplitudes: Odderon input}

In Models I and II, as previously discussed, the scattering amplitude at asymptotic energies is dominated solely by the Pomeron, a
colorless state having the quantum numbers of the vacuum. This particular crossing-even component is precisely the one responsible for
driving the growth of the total cross section with increasing energy.
In the subsequent two models, to be further elaborated upon, we introduce an additional asymptotic term into the
elastic scattering amplitude, characterized by crossing-odd symmetry. This new component, referred to as the Odderon, holds potential
significance in describing soft interactions at high energies.
This leads to the representation of the scattering amplitude in impact parameter space as
\begin{eqnarray}
\chi^{pp}_{\bar{p}p} (s,b) =  \chi_{\Bbb P}(s,b)  + \chi_{+}(s,b)  \pm \chi_{-}(s,b) \pm \xi_{\Bbb O}\chi_{\Bbb O}(s,b) , \nonumber \\
\label{eikonal009}
\end{eqnarray}
where $\chi_{\Bbb O}(s,b)$ represents the Odderon exchange, while $\xi_{\Bbb O}=\pm 1$ is its phase factor. At this stage we have to remember that this phase factor is associated with the positivity property. However, unlike Pomeron, the Odderon is not constrained by positivity requirements.
From a theoretical standpoint, this implies that it is not possible to determine the phase of the Odderon mathematically.
This issue can be succinctly grasped: in the forward direction the physical amplitudes ${\cal F}^{pp}_{\bar{p}p}(s)$ can be written as ${\cal F}^{pp}_{\bar{p}p}(s)=F^{+}(s) \pm F^{-}(s)$. Considering that the only relevant contributions are those arising from the Pomeron and the Odderon exchanges, we can write the symmetric and antisymmetric amplitudes as $F^{+}(s) = R_{\Bbb P}(s) + i I_{\Bbb P}(s)$ and $F^{-}(s) = R_{\Bbb O}(s) + i I_{\Bbb O}(s)$.
From the optical theorem, we have $s \sigma^{pp, \bar{p}p}_{tot}(s) = 4\pi\, \textnormal{Im}\,  {\cal F}^{pp}_{\bar{p}p}(s) > 0$, which implies that
\begin{eqnarray}
\textnormal{Im}\,  {\cal F}^{pp}_{\bar{p}p}(s) =  I_{\Bbb P}(s) \pm I_{\Bbb O}(s) > 0
\end{eqnarray}
and, in turn,
\begin{eqnarray}
I_{\Bbb P}(s) > \left| I_{\Bbb O}(s) \right| .
\end{eqnarray}
As a consequence,
\begin{eqnarray}
I_{\Bbb P}(s) = \frac{s}{2} \left[ \sigma^{pp}_{tot}(s) + \sigma^{\bar{p}p}_{tot}(s)\right] > 0 ,
\end{eqnarray}
while
\begin{eqnarray}
I_{\Bbb O}(s) = \frac{s}{2} \left[ \sigma^{pp}_{tot}(s) - \sigma^{\bar{p}p}_{tot}(s)\right]
\end{eqnarray}
is not bound by the same positivity requirements. Thus, considering that the scattering amplitudes in $b$-space, $\chi_{\Bbb P}(s,b)$ and $\chi_{\Bbb O}(s,b)$, are analytic functions of $s$ with the same cut structure as $F^{+}(s,t)$ and $F^{-}(s,t)$, it becomes evident why the phase of the Odderon term $\chi_{\Bbb O}(s,b)$ remains undefined. Consequently, in all our analyses, we consistently explore both possibilities for $\xi_{\Bbb O}$.

The Born amplitude for the Odderon contribution is represented as
\begin{eqnarray}
{\cal F}_{\Bbb O}(s,t) =  \beta_{\Bbb O}^{2}(t)\, \eta_{\Bbb O}(t) \left( \frac{s}{s_{0}} \right)^{\alpha_{\Bbb O}(t)} ,
\label{odd1}
\end{eqnarray}
where $\eta_{\Bbb O}(t) = -ie^{-i\frac{\pi}{2}\alpha_{\Bbb O}(t)}$.
In the formulation of ``Model III'', we employ an exponential form factor for the proton-Odderon vertex,
\begin{eqnarray}
\beta_{\Bbb O}(t)=\beta_{\Bbb O}(0)\exp \left( \frac{r_{\Bbb O}t}{2} \right) ,
\end{eqnarray}  
with $r_{\Bbb O} = r_{\Bbb P}/2$. This choice is compatible with the constraint $r_{\Bbb O} \leq  r_{\Bbb P}$, which is necessary to avoid the emergence of non-physical effects, such as the appearance of negative cross sections. In this case, the factor 2 relating $r_{\Bbb O}$ and $r_{\Bbb P}$ is not crucial: a change in it is accompanied by a change in the value of $\beta_{\Bbb O}(0)$ that does not substantially alter the fit results.

In the formulation of ``Model IV'', we adopt the power-like form for the proton-Odderon vertex:
\begin{eqnarray}
\beta_{\Bbb O}(t)=\frac{\beta_{\Bbb O}(0)}{(1-t/m_{\rho}^{2})(1-t/a_{\Bbb O})} ,
\label{vertex03}
\end{eqnarray}
where $a_{\Bbb O} = 2 a_{\Bbb P}$. Here, we have a relationship between $a_{\Bbb O}$ and $a_{\Bbb P}$ that must satisfy the constraint $a_{\Bbb O} \geq a_{\Bbb P}$ to avoid non-physical amplitudes when using a power-like form factor. Within Model III (IV), the functional form of the proton-Pomeron vertex bears a resemblance to that of
the Odderon-proton vertex; specifically, it follows an exponential (power-like) form.

To fully characterize the Odderon exchange, we need to determine its trajectory. In line with Regge theory, a compelling option is considering a linear trajectory represented as
\begin{eqnarray}
\alpha_{\Bbb O}(t) = \alpha_{\Bbb O}(0) + \alpha^{\prime}_{\Bbb O} t ,
\label{pomnlin1}
\end{eqnarray}
which satisfies the unitarity constraints $\alpha_{\Bbb O}(0) \leq \alpha_{\Bbb P}(0)$ and
$\alpha^{\prime}_{\Bbb O} \leq \alpha^{\prime}_{\Bbb P}$ \cite{finkeistein01,martynov01,martynov02,martynov03}.
If we write $\alpha_{\Bbb O}(0) = 1 + \delta$, the former constraint can be rewritten as $\delta \leq \epsilon$. A linear trajectory is particularly justified when dealing with an odd amplitude where nonlinear terms from pion loops are negligible. Nevertheless, we can enhance our analysis by incorporating insights from Quantum Chromodynamics (QCD), which has yielded notable and significant advancements in the realm of Odderon studies.

From the standpoint of QCD, when considering the lowest order,
the $C=+1$ amplitude arises from the exchange of two gluons \cite{low01,nussinov01,gunion01,ryskin01,richards01,bopsin01} and the $C=-1$ amplitude from the exchange of three gluons \cite{ewerz01}. These amplitudes exhibit a linear dependence on $s$, meaning that their behavior as $s$ approaches infinity is analogous and corresponds to the presence of a singularity at $j=1$.

Subsequently, extensive theoretical studies have been directed towards uncovering corrections to these results, particularly in higher orders. In this scenario, the leading-log approximation allows for the summation of certain higher-order contributions to physical observables in high-energy particle scattering processes. This approach, where terms of the order $(\alpha_{s} \ln(s))^{n}$ are systematically summed at high energy (large $s$) and small strong coupling $\alpha_{s}$, was widely used in the study of the QCD-Pomeron through the BFKL equation \cite{bfkl01,bfkl02,bfkl03,bfkl04,bfkl05}, which describes the leading logarithmic evolution of gluon ladders in $\ln s$.
As a consequence of this resummation process, the simplistic notion of bare two-gluon exchange gives way to the BFKL Pomeron, which, in an alternative representation, can be seen as the interaction of two reggeized gluons with one another \cite{bopsin01}.

Beyond the BFKL Pomeron, the most elementary entity within perturbative QCD is the exchange involving three interacting reggeized gluons arranged in a symmetric color state. In the framework of the leading-log approximation, the evolution of the three-gluon Odderon exchange as energy increases is governed by the BKP equation \cite{bkp01,bkp02,bkp03}, which involves an iterative application of the BFKL kernel. A bound state solution of the Odderon equation was obtained with the largest intercept found to date, namely $\alpha_{\Bbb O}(0) = 1$ \cite{bartels10}. This solution corresponds to the reggeization of a d-Reggeon in QCD, which interacts with a reggeized gluon. A d-Reggeon is an even-signature color octet that is degenerate with the odd-signature reggeized gluon.
Based on these QCD findings, we adopt in this work the simplest conceivable form for the Odderon trajectory:
\begin{eqnarray}
\alpha_{\Bbb O}(t) = 1 .
\label{pomnlin2}
\end{eqnarray}

\begin{table*}
\centering
\caption{The Pomeron and secondary Reggeons parameters values obtained in global fits to Ensembles A and T after the eikonal unitarization.}
\begin{ruledtabular}
\begin{tabular}{ccccc}
 & \multicolumn{2}{c}{Ensemble A} & \multicolumn{2}{c}{Ensemble T} \\
\cline{2-3} \cline{4-5} 
 & Model I & Model II & Model I & Model II  \\
\hline
$\epsilon$ & 0.1014$\pm$0.0033 & 0.1112$\pm$0.0013 & 0.1248$\pm$0.0027 & 0.1336$\pm$0.0023 \\
$\alpha^{\prime}_{I\!\!P}$ (GeV$^{-2}$) & 0.2938$\pm$0.0022 & 0.1148$\pm$0.0076 & 0.56 $\times$ 10$^{-9}$ $\pm$0.11 & 0.009$\pm$0.040   \\
$\beta_{\Bbb P}(0)$ & 2.154$\pm$0.063 & 1.999$\pm$0.023 & 1.814$\pm$0.043 & 1.742$\pm$0.028  \\
$r_{\Bbb P}$ (GeV$^{-2}$) & 2.375$\pm$0.019 & --- & 7.448$\pm$0.087 & ---  \\
$a_{\Bbb P}$ (GeV$^{-2}$) & --- & 0.829$\pm$0.081 & --- & 0.499$\pm$0.084  \\
$\eta_{+}$ & 0.360$\pm$0.048 & 0.344$\pm$0.030 & 0.286$\pm$0.025 & 0.262$\pm$0.015  \\
$\beta_{+}(0)$ & 4.56$\pm$0.47 & 4.37$\pm$0.34 & 4.02$\pm$0.21 & 3.93$\pm$0.14  \\
$\eta_{-}$ & 0.556$\pm$0.010 & 0.550$\pm$0.089 & 0.536$\pm$0.067 & 0.530$\pm$0.064  \\
$\beta_{-}(0)$ & 3.68$\pm$0.16 & 3.55$\pm$0.67 & 3.41$\pm$0.49 & 3.39$\pm$0.46  \\
\hline
$\nu$ & 226 & 226 & 350 & 350  \\
\hline
$\chi^{2}/\nu$ & 0.86 & 0.83 & 0.74  & 0.65  \\
\end{tabular}
\end{ruledtabular}
\label{tab001}
\end{table*}

\begin{table*}
\centering
\caption{The Pomeron and secondary Reggeons parameters values obtained in global fits to Ensembles A and T after the $U$-matrix unitarization.}
\begin{ruledtabular}
\begin{tabular}{ccccc}
 & \multicolumn{2}{c}{Ensemble A} & \multicolumn{2}{c}{Ensemble T} \\
\cline{2-3} \cline{4-5} 
 & Model I & Model II  & Model I & Model II  \\
\hline
$\epsilon$ & 0.0911$\pm$0.0037 & 0.0981$\pm$0.0029 & 0.1129$\pm$0.0048 & 0.1150$\pm$0.0070 \\
$\alpha^{\prime}_{I\!\!P}$ (GeV$^{-2}$) & 0.4425$\pm$0.0085 & 0.2728$\pm$0.0089 & 0.05$\pm$0.14 & 0.10$\pm$0.12   \\
$\beta_{\Bbb P}(0)$ & 2.271$\pm$0.075 & 2.140$\pm$0.056 & 1.926$\pm$0.085 & 1.92$\pm$0.11  \\
$r_{\Bbb P}$ (GeV$^{-2}$) & 0.1051$\pm$0.0061 & --- & 7.2$\pm$2.8 & ---  \\
$a_{\Bbb P}$ (GeV$^{-2}$) & --- & 40$\pm$20 & --- & 0.62$\pm$0.49  \\
$\eta_{+}$ & 0.356$\pm$0.057 & 0.369$\pm$0.049 & 0.325$\pm$0.050 & 0.314$\pm$0.053  \\
$\beta_{+}(0)$ & 4.71$\pm$0.65 & 4.51$\pm$0.48 & 4.18$\pm$0.43 & 4.14$\pm$0.44  \\
$\eta_{-}$ & 0.551$\pm$0.098 & 0.551$\pm$0.043 & 0.545$\pm$0.074 & 0.542$\pm$0.075  \\
$\beta_{-}(0)$ & 3.59$\pm$0.74 & 3.54$\pm$0.34 & 3.43$\pm$0.54 & 3.43$\pm$0.54  \\
\hline
$\nu$ & 226 & 226 & 350 & 350  \\
\hline
$\chi^{2}/\nu$ & 0.85 & 0.86 & 0.71  & 0.64  \\
\end{tabular}
\end{ruledtabular}
\label{tab002}
\end{table*}

\section{Results and conclusions}

\begin{figure*}\label{fig001}
\begin{center}
\includegraphics[height=.70\textheight]{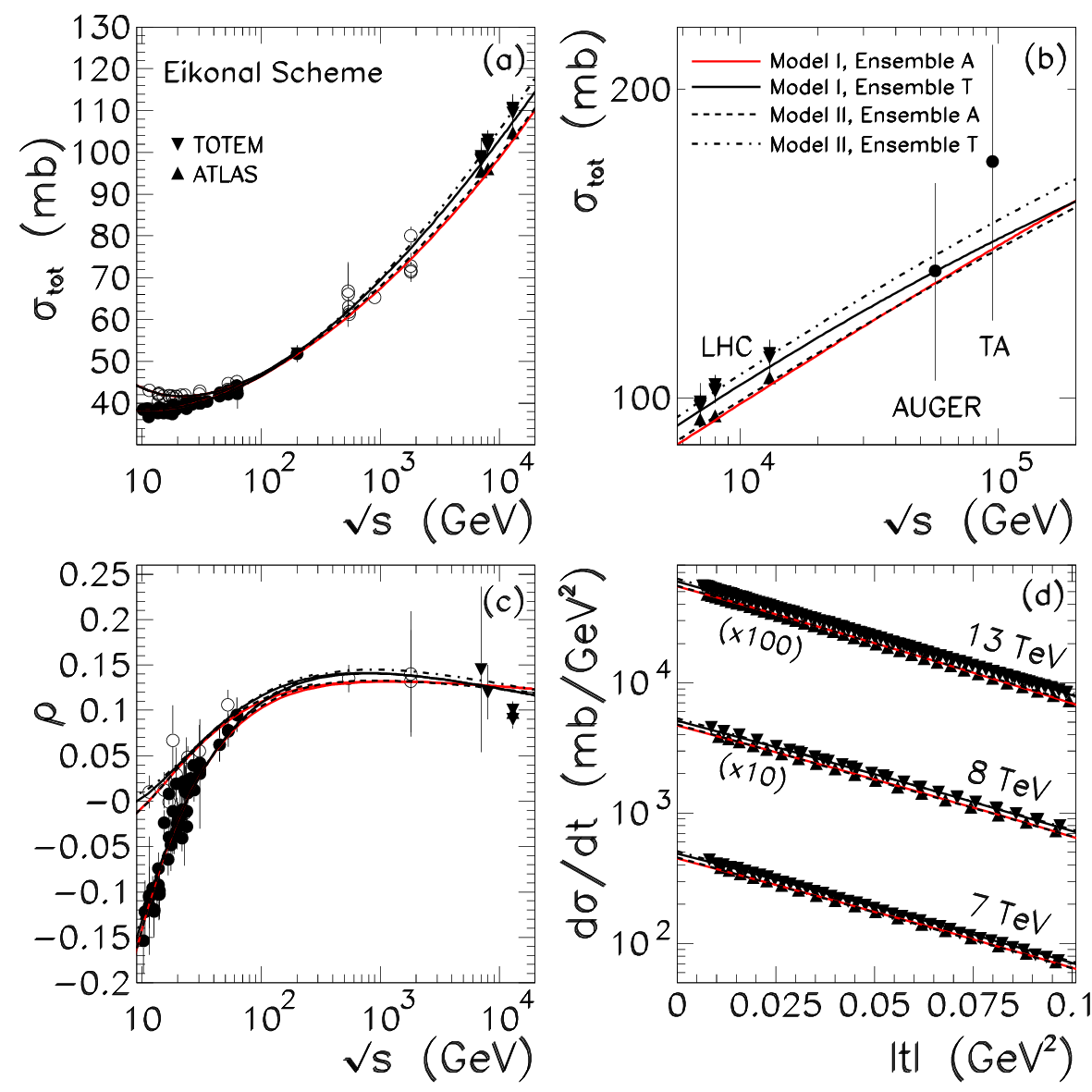}
\caption{Total cross section, $\rho$ parameter, and differential cross section for $pp$ ($\bullet$, $\blacktriangle$, $\blacktriangledown$) and $\bar{p}p$ ($\circ$) channels. Results obtained using eikonal unitarization.}
\end{center}
\end{figure*}

\begin{figure*}\label{fig002}
\begin{center}
\includegraphics[height=.70\textheight]{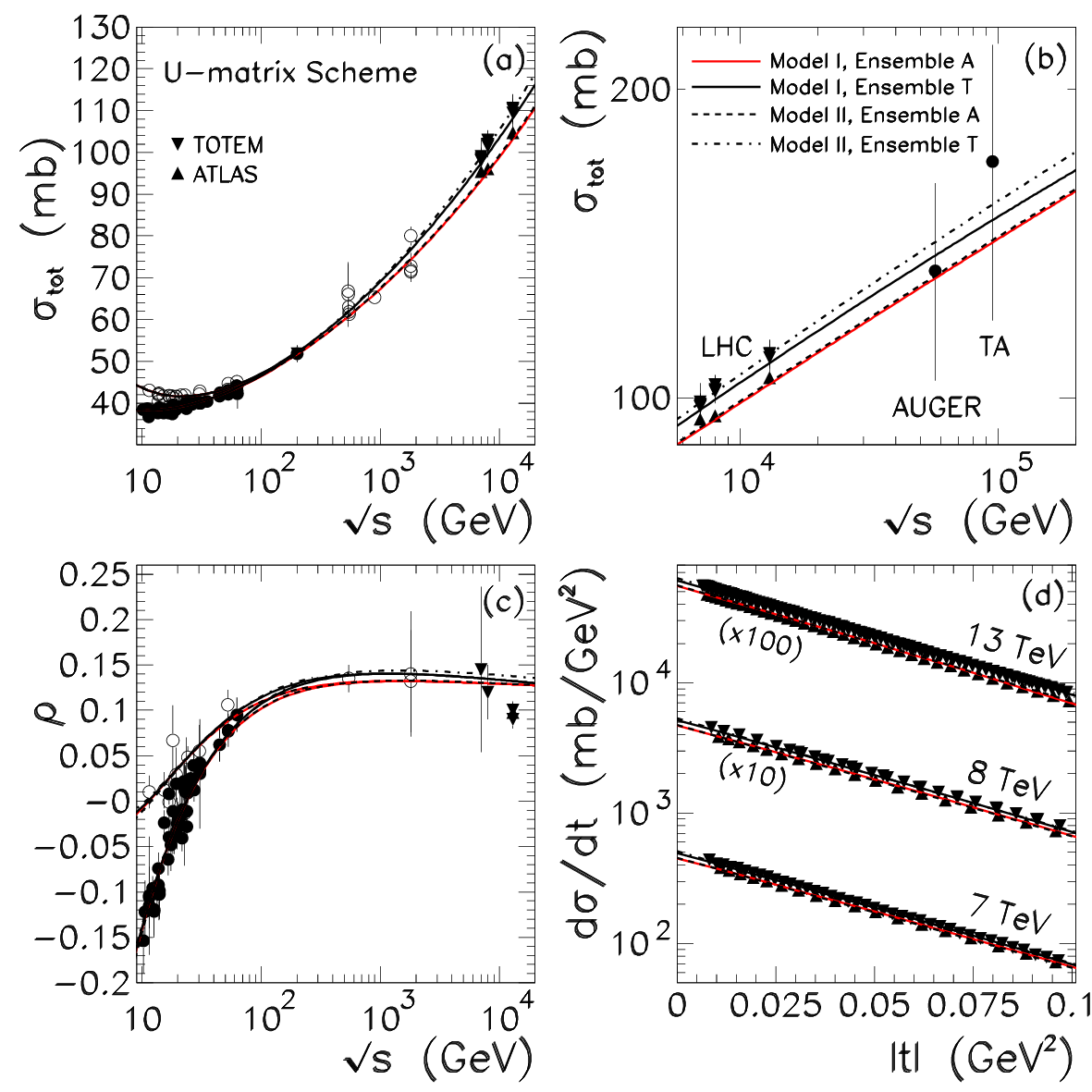}
\caption{Total cross section, $\rho$ parameter, and differential cross section for $pp$ ($\bullet$, $\blacktriangle$, $\blacktriangledown$) and $\bar{p}p$ ($\circ$) channels. Results obtained using $U$-matrix unitarization.}
\end{center}
\end{figure*}

The LHC has released exceptionally precise measurements of diffractive processes, offering a unique constraint on the behavior of scattering amplitudes at high energies. These measurements, particularly the total and differential cross sections obtained from ATLAS and TOTEM Collaborations \cite{antchev001,antchev002,TOTEM001,TOTEM005,TOTEM008,TOTEM010,atlas001,atlas002,ATLAS01}, enable us to determine the Pomeron and Odderon parameters accurately.

However, these experimental results unveil a noteworthy tension between the TOTEM and ATLAS measurements \cite{petrov01,petrov02}. For instance, when comparing the TOTEM result for $\sigma_{tot}^{pp}$ at $\sqrt{s}=7$ TeV, $\sigma_{tot}^{pp}=98.58\pm 2.23$ \cite{antchev001}, with the most precise value recorded by ATLAS at the same energy, $\sigma_{tot}^{pp}=95.35\pm 1.36$ \cite{atlas001}, the discrepancy between these values, assuming uncorrelated uncertainties, corresponds to 1.4 $\sigma$. Similarly, comparing the ATLAS result for the total cross section at $\sqrt{s}=8$ TeV, $\sigma_{tot}^{pp}=96.07\pm 0.92$ \cite{atlas002}, with the lowest TOTEM measurement at the same center-of-mass energy, $\sigma_{tot}^{pp}=101.5\pm 2.1$ \cite{antchev002}, reveals an even more pronounced difference: 2.6 $\sigma$.

This significant disagreement suggests diverse scenarios for the rise of the total cross section and, consequently, for the parameters of the Pomeron and the Odderon.

In order to systematically explore the tension between TOTEM and ATLAS results, we perform global fits to $pp$ and $\bar{p}p$ forward scattering data above $\sqrt{s} = 10$ GeV and to $pp$ differential cross-section data while considering two distinct datasets, one with TOTEM measurements and the other with ATLAS measurements. Precisely, we fit to the total cross section, $\sigma_{tot}^{pp,\bar{p}p}$, to the ratio of the real to the imaginary part of the scattering amplitude, $\rho^{pp,\bar{p}p}$, and to the elastic differential cross section, $d\sigma^{pp}/dt$, at 7, 8, and 13 TeV with $|t| \leq 0.1$ GeV$^{2}$, which are segregated into two distinct datasets: one featuring TOTEM measurements and the other containing ATLAS measurements. We use $\sigma_{tot}$ and $\rho$ data with $\sqrt{s}\geq 10$ GeV compiled and analyzed by the Particle Data Group \cite{pdg001}, and $d\sigma/dt$ data at LHC from the TOTEM \cite{antchev001,antchev002,TOTEM001,TOTEM005,TOTEM008,TOTEM010} and ATLAS \cite{atlas001,atlas002,ATLAS01} Collaborations. The statistical and systematic errors were added in quadrature for all the data.

The procedure of separating discrepant data measured at the same center-of-mass energy and, in this way, generating two different data ensembles is statistically robust and has been previously employed to investigate discrepancies in cosmic-ray data and how they affect the predictions for $pp$ total cross sections at very high energies \cite{luna011}. This same procedure has also been applied to analyze the tension between CDF and E710/E811 Tevatron data and their effects on determining extrema bounds for the Pomeron intercept \cite{luna012a,luna012b,luna012c}. 

The two data ensembles can be defined as follows:

{\bf Ensemble A}: $\sigma_{tot}^{pp,\bar{p}p}$ data + $\rho^{pp,\bar{p}p}$ data + ATLAS data on $\frac{d\sigma}{dt}$ at 7, 8, and 13 TeV;

{\bf Ensemble T}: $\sigma_{tot}^{pp,\bar{p}p}$ data + $\rho^{pp,\bar{p}p}$ data + TOTEM data on $\frac{d\sigma}{dt}$ at 7, 8, and 13 TeV.

The Ensemble A (T) includes $\sigma_{tot}^{pp}$ and $\rho^{pp}$ data measured by the ATLAS (TOTEM) collaboration. In the case of the $d\sigma/dt$ data, we establish a lower limit for $|t|$, namely $|t|_{min} \sim 10|t|_{int}$, where $|t|_{int}$ designates the point at which the interference between the Coulomb and hadronic amplitudes is maximal. The value of $|t|_{int}$ can be conveniently determined using the empirical relationship  $|t|_{int}=0.071/\sigma_{tot}$ \cite{blockcahn}.

With our data sets clearly defined, we turn to phenomenology and carry out global fits to the two distinct ensembles using a $\chi^{2}$ fitting procedure, where $\chi^{2}_{min}$ follows a $\chi^{2}$ distribution with $\nu$ degrees of freedom. These global fittings for the experimental data incorporate a $\chi^{2}$ interval, which, in the case of normal errors, corresponds to the region encompassing 90\% of the probability within the $\chi^{2}$ hypersurface. This corresponds to $\chi^{2}-\chi^{2}_{min}=13.36$ and $14.68$ for eight and nine free parameters, respectively. Consistent with our approach of minimizing the number of free parameters, we fix the slopes of the secondary-Reggeon linear trajectories, $\alpha^{\prime}_{+}$ and $\alpha^{\prime}_{-}$, at 0.9 GeV$^{-1}$. These values align with the typical values observed in Chew-Frautschi plots. Additionally, we set the slopes related to the form factors of the secondary Reggeons to $r_{+} = r_{-} = 4.0$ GeV$^{-2}$. These parameters exhibit minimal statistical correlation with the Pomeron (and Odderon) parameters, and their fixed values are in agreement with those obtained in previous studies \cite{goulianos001,kmr002,kmr003}.

\begin{table*}
\centering
\caption{The Pomeron, Odderon and secondary Reggeons parameters values obtained in global fits to Ensembles A and T after the eikonal unitarization. We show the results with $\xi_{\Bbb O} = -1$.}
\begin{ruledtabular}
\begin{tabular}{ccccc}
 & \multicolumn{2}{c}{Ensemble A} & \multicolumn{2}{c}{Ensemble T} \\
\cline{2-3} \cline{4-5} 
 & Model III & Model IV & Model III & Model IV  \\
\hline
$\epsilon$ & 0.1017$\pm$0.0043 & 0.1043$\pm$0.0026 & 0.1247$\pm$0.0048 & 0.1335$\pm$0.0041 \\
$\alpha^{\prime}_{I\!\!P}$ (GeV$^{-2}$) & 0.283$\pm$0.036 & 0.242$\pm$0.012 & 0.94 $\times$ 10$^{-4}$ $\pm$0.059 & 0.01$\pm$0.11   \\
$\beta_{\Bbb P}(0)$ & 2.146$\pm$0.083 & 2.116$\pm$0.011 & 1.815$\pm$0.080 & 1.744$\pm$0.035  \\
$r_{\Bbb P}$ (GeV$^{-2}$) & 2.58$\pm$0.68 & --- & 7.45$\pm$0.13 & ---  \\
$a_{\Bbb P}$ (GeV$^{-2}$) & --- & 31$\pm$11 & --- & 0.50$\pm$0.16  \\
$\beta_{\Bbb O}(0)$ & 0.47$\pm$0.24 & 0.40$\pm$0.17 & 0.31$\pm$0.24 & 0.27$\pm$0.20  \\
$\eta_{+}$ & 0.359$\pm$0.055 & 0.353$\pm$0.020 & 0.285$\pm$0.051 & 0.261$\pm$0.013  \\
$\beta_{+}(0)$ & 4.52$\pm$0.54 & 4.47$\pm$0.29 & 4.00$\pm$0.38 & 3.91$\pm$0.16  \\
$\eta_{-}$ & 0.4823$\pm$0.0019 & 0.482$\pm$0.077 & 0.490$\pm$0.030 & 0.489$\pm$0.077  \\
$\beta_{-}(0)$ & 3.20$\pm$0.13 & 3.19$\pm$0.50 & 3.14$\pm$0.22 & 3.15$\pm$0.50  \\
\hline
$\nu$ & 225 & 225 & 349 & 349  \\
\hline
$\chi^{2}/\nu$ & 0.84 & 0.80 & 0.73  & 0.65  \\
\end{tabular}
\end{ruledtabular}
\label{tab003}
\end{table*}

\begin{table*}
\centering
\caption{The Pomeron, Odderon and secondary Reggeons parameters values obtained in global fits to Ensembles A and T after the $U$-matrix unitarization. We show the results with $\xi_{\Bbb O} = -1$.}
\begin{ruledtabular}
\begin{tabular}{ccccc}
 & \multicolumn{2}{c}{Ensemble A} & \multicolumn{2}{c}{Ensemble T} \\
\cline{2-3} \cline{4-5} 
 & Model III & Model IV & Model III & Model IV  \\
\hline
$\epsilon$ & 0.0938$\pm$0.0045 & 0.0978$\pm$0.0047 & 0.1115$\pm$0.0035 & 0.1148$\pm$0.0060 \\
$\alpha^{\prime}_{I\!\!P}$ (GeV$^{-2}$) & 0.364$\pm$0.029 & 0.273$\pm$0.031 & 0.10$\pm$0.15 & 0.106$\pm$0.098   \\
$\beta_{\Bbb P}(0)$ & 2.215$\pm$0.075 & 2.146$\pm$0.066 & 1.951$\pm$0.063 & 1.919$\pm$0.093  \\
$r_{\Bbb P}$ (GeV$^{-2}$) & 1.57$\pm$0.58 & --- & 6.2$\pm$3.0 & ---  \\
$a_{\Bbb P}$ (GeV$^{-2}$) & --- & 40$\pm$24 & --- & 0.63$\pm$0.41  \\
$\beta_{\Bbb O}(0)$ & 0.44$\pm$0.20 & 0.23$\pm$0.15 & 0.32$\pm$0.18 & 0.27$\pm$0.18  \\
$\eta_{+}$ & 0.374$\pm$0.031 & 0.369$\pm$0.026 & 0.327$\pm$0.071 & 0.313$\pm$0.046  \\
$\beta_{+}(0)$ & 4.62$\pm$0.50 & 4.49$\pm$0.64 & 4.18$\pm$0.72 & 4.12$\pm$0.38  \\
$\eta_{-}$ & 0.490$\pm$0.047 & 0.48$\pm$0.33 & 0.49$\pm$0.21 & 0.50$\pm$0.12  \\
$\beta_{-}(0)$ & 3.18$\pm$0.18 & 3.08$\pm$0.79 & 3.11$\pm$0.42 & 3.17$\pm$0.71  \\
\hline
$\nu$ & 225 & 225 & 349 & 349  \\
\hline
$\chi^{2}/\nu$ & 0.83 & 0.84 & 0.71  & 0.64  \\
\end{tabular}
\end{ruledtabular}
\label{tab004}
\end{table*}

\begin{figure*}\label{fig003}
\begin{center}
\includegraphics[height=.70\textheight]{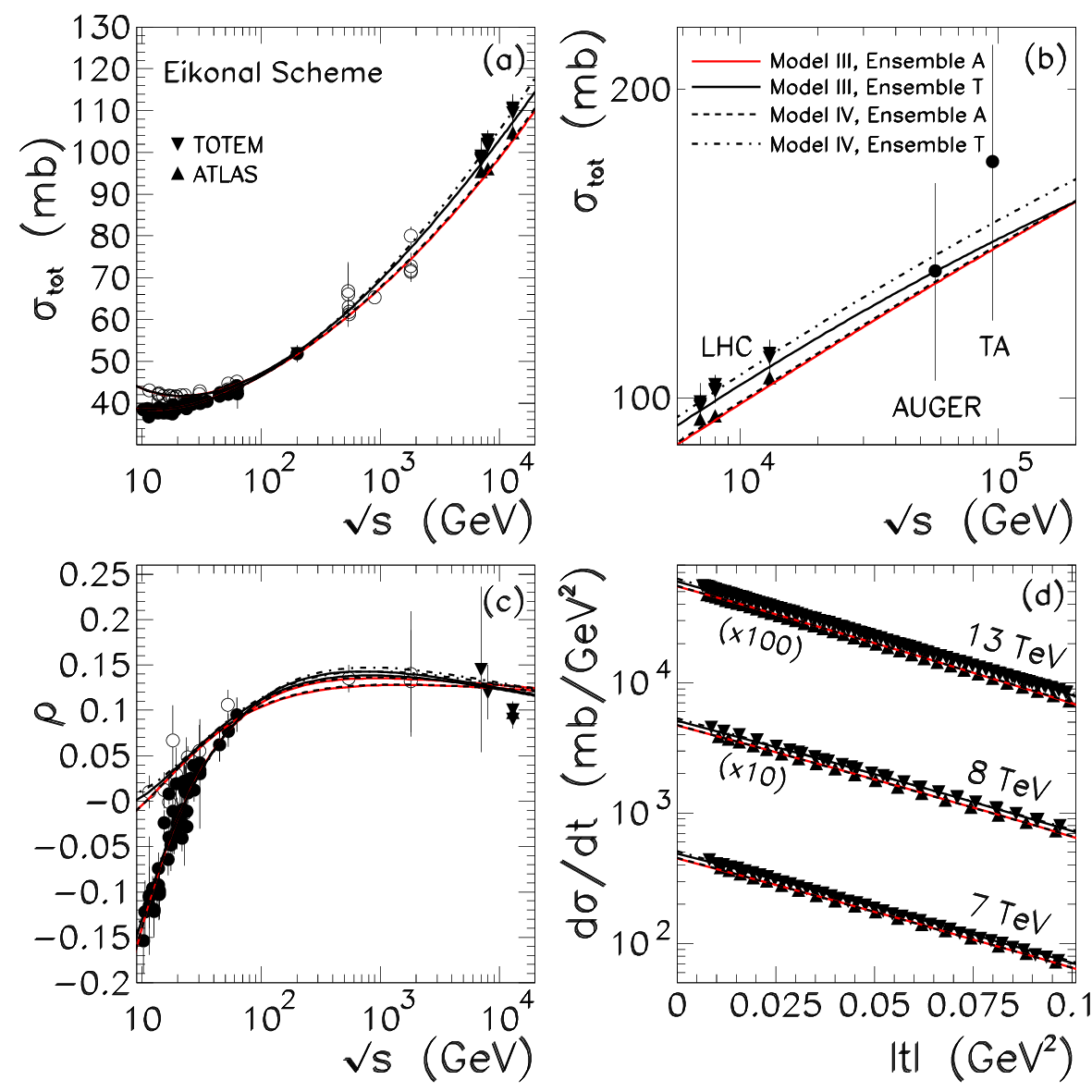}
\caption{Total cross section, $\rho$ parameter, and differential cross section for $pp$ ($\bullet$, $\blacktriangle$, $\blacktriangledown$) and $\bar{p}p$ ($\circ$) channels. Results obtained using eikonal unitarization.}
\end{center}
\end{figure*}

\begin{figure*}\label{fig004}
\begin{center}
\includegraphics[height=.70\textheight]{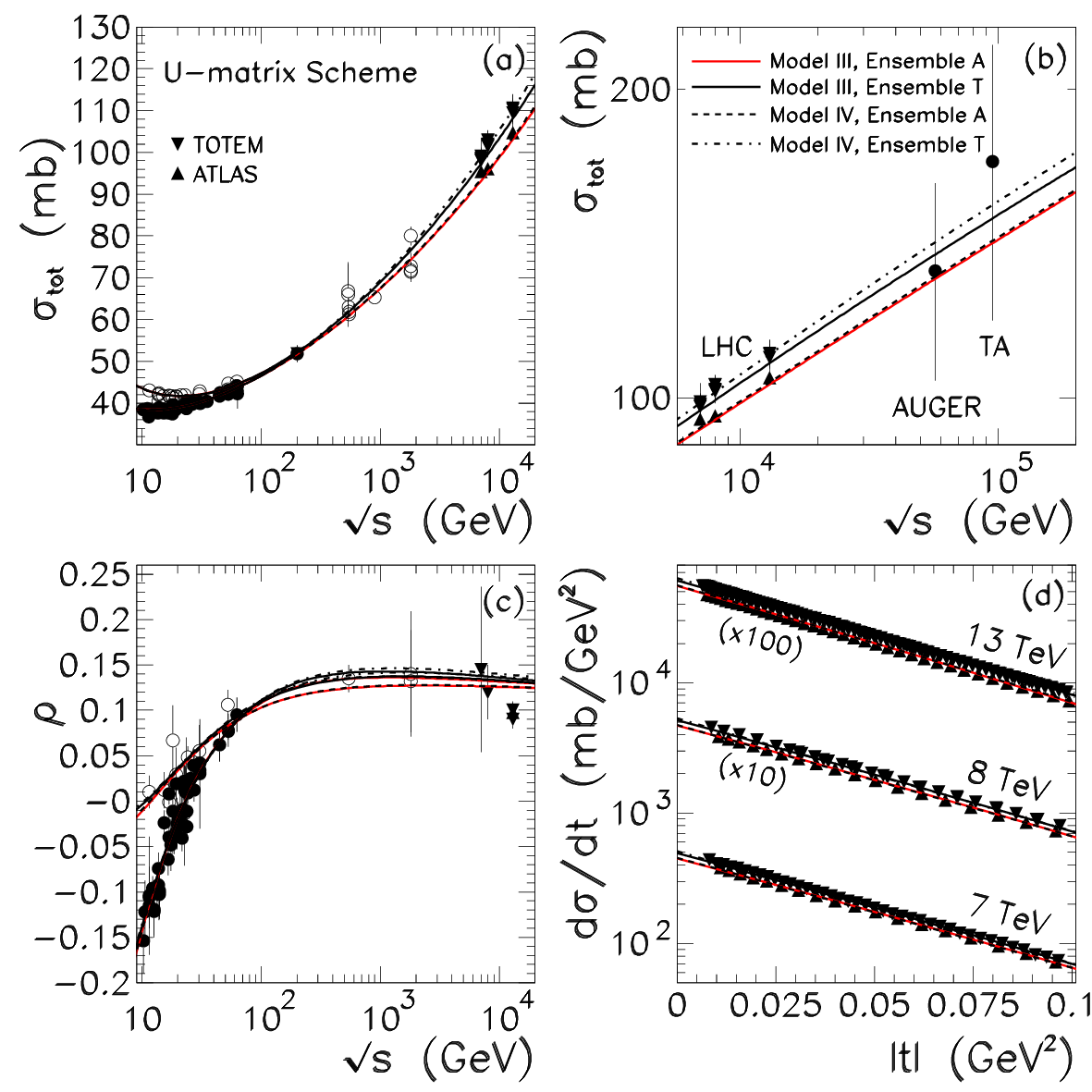}
\caption{Total cross section, $\rho$ parameter, and differential cross section for $pp$ ($\bullet$, $\blacktriangle$, $\blacktriangledown$) and $\bar{p}p$ ($\circ$) channels. Results obtained using $U$-matrix unitarization.}
\end{center}
\end{figure*}

\subsection{Pomeron Analysis}

The Pomeron and the secondary Reggeon parameters determined by the global fits to Ensembles A and T, as derived from Models I and II, within the context of the eikonal and $U$-matrix unitarization schemes, are presented in Tables \ref{tab001} and \ref{tab002}, respectively. The results of these fits are shown in Figures 1 (eikonal scheme) and 2 ($U$-matrix scheme). We first observe that the parameters related to secondary Reggeons are not sensitive to the unitarization scheme since their values are compatible with each other considering the associated uncertainties.

The same is not observed concerning the Pomeron intercept: the values of the $\epsilon$ parameter are systematically higher when determined through the eikonal unitarization. Specifically, when switching from the $U$-matrix to the eikonal scheme, particularly in Ensemble A, the results show a significant increase of 11.3\% in the $\epsilon$ parameter for Models I and II. Similarly, in Ensemble T, this change leads to an increase of 10.5\% in $\epsilon$ for Model I and 16.2\% for Model II.

An opposite situation occurs in the case of the $\beta_{\Bbb P}(0)$ parameter: since the parameter $\epsilon$ controls the energy dependence of the total cross sections, we expect that an increase in $\epsilon$ will be compensated by a decrease in the value of the coupling $\beta_{\Bbb P}(0)$, which, in fact, happens: all values of $\beta_{\Bbb P}(0)$ determined in the $U$-matrix scheme are higher than their counterparts in the eikonal scheme. Specifically, for Ensemble A and Model I (Model II), the increase is about 5.4\% (7.1\%); for Ensemble T and Model I (Model II), the increase stands at around 6.2\% (10.2\%).

For the $\alpha^{\prime}_{I\!\!P}$ parameter, its values are always more significant in the $U$-matrix scheme than in the eikonal one. For Ensemble A, when compared to the values in the eikonal scheme, the increase of $\alpha^{\prime}_{I\!\!P}$ is about 51\% in the case of Model I and increases by a factor of 2.4 in the case of Model II.
Interestingly enough, for Ensemble T, in both models and unitarization schemes all values of $\alpha^{\prime}_{I\!\!P}$ are consistent with zero, i.e., the trajectory of the Pomeron is consistent with the simple form $\alpha_{I\!\!P}(t)= 1 + \epsilon$ (a fixed pole). It is worth remembering that this behavior is consistent with that obtained in screened Regge models, where very small values of $\alpha^{\prime}_{I\!\!P}$ are obtained \cite{kmr002,kmr003,kmrepj01,kmrepj02,lev01,lev02}. In Gribov Reggeon calculus, the smallness of $\alpha^{\prime}_{I\!\!P}$ indicates that, in principle, the soft Pomeron can be treated perturbatively \cite{gribov01,gribov02,gribov03,baker01}, raising the possibility of formulating a fundamental theory for soft processes based exclusively on QCD.

When examining the parameters related to the proton-Pomeron vertices, we transition from a situation in which the parameters are not sensitive to the unitarization scheme used (as observed in Ensemble T) to another situation where a notable variation in the values of $r_{\Bbb P}$ and $a_{\Bbb P}$ occurs when transitioning between schemes (as evident in Ensemble A). In the specific case of Model II, Ensemble A, the value of $a_{\Bbb P}$ determined in the $U$-matrix scheme is more significant than that obtained in the eikonal scheme by a factor of around 48.

Figures 1 and 2 describe $pp$ and $\bar{p}p$ data for Models I and II. Specifically, their respective parts (a) illustrate the total cross-section, (b) extend the range in $\sqrt{s}$ of the part (a), (c) highlight the $\rho$ parameter (the ratio of real to imaginary parts of the scattering amplitude) and (d) focus on the differential cross section. We have included in parts (b), for comparison purposes, estimates of $pp$ total cross section obtained from cosmic ray experiments, namely, the AUGER result at $\sqrt{s} = 57$ TeV \cite{auger} and the Telescope Array result at $\sqrt{s} = 95$ TeV \cite{TA}.

In the eikonal scheme, we notice a distinctive feature for Model I – crossing of the total cross sections at extremely high energies (Figure 1b). This crossing is mirrored in the behavior of the $\rho$ parameter (Figure 1c). Consequently, while the cross sections associated with Ensemble T consistently surpass those obtained from Ensemble A at LHC energies, a reversal occurs at energies beyond those of cosmic rays. This inversion is not present when unitarization is accomplished using the $U$-matrix scheme (Figure 2b). At $\sqrt{s} \sim $ 100 TeV, the highest (smallest) value for the total cross section is achieved when employing the combination of Model II with Ensemble T (Model II with Ensemble A).

We see that in the analyses taking into account only the Pomeron's presence, it is not possible to assert the superiority of one unitarization scheme over the other based only on the $\chi^{2}/\nu$ values. The most significant fluctuation in the $\chi^{2}/\nu$ value after a transition between unitarization schemes is approximately 4\%. Moreover, the smallest values of $\chi^{2}/\nu$ intermittently manifest in analyses using both the eikonal and $U$-matrix schemes. 
Nevertheless, compared with the experimental values of $\rho$ at 13 TeV measured by the ATLAS and TOTEM teams, namely $\rho^{pp} = 0.09\pm0.01$, $\rho^{pp} = 0.10\pm0.01$ (TOTEM extractions exploiting the Coulomb-nuclear interference) \cite{TOTEM008}, and $\rho^{pp} = 0.098\pm0.011$ (ATLAS extraction from a fit to $d\sigma/dt$ using the optical theorem) \cite{ATLAS01}, the eikonal scheme exhibits a slight advantage over the U-matrix scheme. Despite not being precisely aligned with the experimental central values, the eikonal scheme predictions are closer to the measurements obtained at the LHC. Specifically, in the eikonal scheme, for Model I with Ensemble A, Model I with Ensemble T, Model II with Ensemble A, and Model II with Ensemble T, the predictions for the $\rho$ parameter at 13 TeV are 0.125, 0.121, 0.122, and 0.127, respectively. In contrast, the corresponding predictions in the $U$-matrix scheme are 0.129, 0.132, 0.129, and 0.138, respectively. These findings suggest, therefore, that an analysis solely relying on the Pomeron is insufficient for providing an accurate description of the $\rho$ data measured at $\sqrt{s} = 13$ TeV.

\subsection{Pomeron $\oplus$ Odderon Analysis}

We proceed with our analyses by first considering the presence of an Odderon with a phase factor $\xi_{\Bbb O}= -1$. Tables \ref{tab003} and \ref{tab004} present the Pomeron, the Odderon, and the secondary Reggeon parameters determined by the global fits to Ensembles A and T, as derived from Models III and IV. These fits were performed within the frameworks of the eikonal and $U$-matrix unitarization schemes. The results of these fits are depicted in Figures 3 (eikonal scheme) and 4 ($U$-matrix scheme). These Figures have the same layout as the Figures 1 and 2. It is evident that for Ensemble T, the inclusion of the Odderon did not result in a decrease in the values of $\chi^{2}/\nu$ when compared with the values obtained in the analysis without the Odderon. Specifically, the values of $\chi^{2}/\nu$ remain the same in the $U$-matrix scheme for Models III and IV, as well as in the eikonal scheme for Model IV. A variation is observed only in the eikonal scheme and Model III case, but the reduction is only 1.4\% compared to the result for Model I. For Ensemble A, we see that the introduction of the Odderon reduces the value of $\chi^{2}/\nu$ across all scenarios. In the eikonal scheme for Model III, the decrease is around 3.6\%, while in the other three remaining cases, the decrease hovers around 2.3\%. However, from a statistical standpoint, these variations are not significant enough to conclusively assert the indispensability of Odderon's presence. A more thorough examination of Odderon's significance in these analyses can be achieved by assessing its impact on describing the $\rho$ parameter at high energies, a discussion we will delve into later.

Once again, the parameters associated with secondary Reggeons are not sensitive to the unitarization scheme, as their values remain compatible, accounting for the associated uncertainties. In the same way as in the analysis without Odderon, the values of the $\epsilon$ parameter are systematically higher when determined through the eikonal unitarization. More precisely, when switching from the $U$-matrix to the eikonal scheme, in Ensemble A, the results show an increase of 8.4\% for Model III and 6.6\% for Model IV; in Ensemble T, this change results in an escalation of 11.8\% for Model III and 16.3\% for Model IV.

As expected, the increase in $\epsilon$ parameter in the transition from $U$-matrix to eikonal scheme is compensated by the decrease in the value of the coupling $\beta_{\Bbb P}(0)$: for Ensemble A and Model III (Model IV), the increase is about 3.2\% (1.4\%); for Ensemble T and Model III (Model IV), the rise amounts to around 7.5\% (10.0\%). In this scenario, the average increase for Ensemble A is relatively minor compared to the scenario without the Odderon. Conversely, for Ensemble T, the increments are nearly identical to the case when considering only the Pomeron.

Regarding the $\alpha^{\prime}_{I\!\!P}$ parameter, a similar pattern emerges as observed in the case without the Odderon: its magnitudes consistently exceed those in the eikonal scheme within the $U$-matrix unitarization. In Ensemble A, compared to the values in the eikonal scheme, there is an approximately 29\% increase in $\alpha^{\prime}_{I\!\!P}$ for Model III and 12.8\% for Model IV. Notably, for Ensemble T, in both models and unitarization schemes, all values of $\alpha^{\prime}_{I\!\!P}$ align with zero, implying that the trajectory of the Pomeron is consistent with the form $\alpha_{I\!\!P}(t)= 1 + \epsilon$. Therefore, even after the introduction of the Odderon, for Ensemble T, the analysis continues to favor the existence of a fixed Pomeron pole.

After analyzing the parameters associated with the proton-Pomeron vertices, it is clear that the inclusion of the Odderon has no noticeable effect on the properties of $r_{\Bbb P}$ and $a_{\Bbb P}$. The values obtained using the eikonal method are currently consistent with those obtained using the $U$-matrix method, considering the associated uncertainties. It is noteworthy to recall that the Pomeron parameters $r_{\Bbb P}$ and $a_{\Bbb P}$ maintain a proportional relationship (by a factor of 2) to the Odderon parameters $r_{\Bbb O}$ and $a_{\Bbb O}$.

A more discernible indication of Odderon's presence emerges when scrutinizing the values of the $\beta_{\Bbb O}(0)$ coupling. Notably, all values of this coupling exhibit non-zero magnitudes across all analyses. Despite the uncertainties associated with it consistently surpassing those of the Pomeron coupling $\beta_{\Bbb P}(0)$, none of the cases yield values consistent with zero. Although there may be slight variations in the values of $\beta_{\Bbb O}(0)$ when the proportionality factor between $r_{\Bbb P}$ and $r_{\Bbb O}$ and between $a_{\Bbb P}$ and $a_{\Bbb O}$ deviates from 2 (we tested values between 2 and 5), the observed changes are negligible. From the Eikonal unitarization scheme, the ratios of Odderon to Pomeron exchange couplings denoted as $\beta^{2}_{\mathbb{O}}(0)/\beta^{2}_{\mathbb{P}}(0)$, exhibit values of 0.048 and 0.036 for Models III and IV, respectively, within Ensemble A. Meanwhile, for Ensemble T, these ratios are 0.029 and 0.024 for Models III and IV, respectively. Similarly, under the $U$-matrix unitarization scheme, the corresponding ratios for Ensemble A are 0.039 and 0.011 for Models III and IV. For Ensemble T, they are 0.027 and 0.020 for Models III and IV, respectively.

The presence of the Odderon immediately impacts the behavior of total cross sections, particularly generating different growth patterns for $\sigma_{tot}^{pp}(s)$ and $\sigma_{tot}^{\bar{p}p}(s)$ at high energies. If a scattering amplitude is dominated asymptotically by only crossing-even terms, we have $\left| \Delta \sigma \right| = \left| \sigma_{tot}^{\bar{p}p} - \sigma_{tot}^{pp} \right| \to$ 0 at $s\to \infty$. However, supposing there is an asymptotic non-zero crossing-odd term
${\cal A}^{-}(s,t)$ in the scattering amplitude, it is possible to demonstrate via Froissart-Martin theorem that the difference $\left| \Delta \sigma \right|$ can be at most $\left| \Delta \sigma \right| = k \ln s $ in the limit $s\to \infty$, where $k$ is a constant \cite{delta1,delta2}. Furthermore, when ${\cal A}^{-}(s,t)$ does not disappear at high energies, the cross sections also obey the general Pomeranchuk theorem $\left[ \sigma_{tot}^{\bar{p}p}(s)/\sigma_{tot}^{pp}(s) \right]_{s\to \infty} \to  1$ \cite{pmrchuk1,pmrchuk2,pmrchuk3}.
Regrettably, Figures 3(a) and 4(a) do not provide a precise observation of this phenomenon, as the $\sigma_{tot}^{pp}(s)$ and $\sigma_{tot}^{\bar{p}p}(s)$ curves closely overlap in all four cases. Nevertheless, the distinctive behavior unveils itself in the behavior of the $\rho$ parameter, as shown in Figures 3(c) and 4(c).
The reason for this is relatively simple: the parameter $\rho(s)$ can be derived from $\sigma_{tot}(s)$ using derivative dispersion relations \cite{lunphys01,lunphys02}. In essence, the behavior of $\rho(s)$ is intricately tied to changes in the curvature of $\sigma_{tot}(s)$. Consequently, even subtle differences between the growth of $\sigma^{pp}_{tot}(s)$ and $\sigma^{\bar{p}p}_{tot}(s)$ become discernible when expressed through the lens of $\rho$.
Thus, in analyses concerning the Odderon, the predicted value of the $\rho$ parameter at $\sqrt{s}=13$ TeV holds significant importance. As already mentioned, the observables $\sigma_{tot}$ and $\rho$ are intricately connected through dispersion relations derived from fundamental unitarity and analyticity principles applied to scattering amplitudes. However, the importance of the parameter $\rho$ goes beyond just monitoring the high-energy progression of the total hadronic cross section; it also provides insights into the fundamental structure of the elastic-scattering amplitude.
From the QCD viewpoint, the potential inclusion of an additional three-gluon colorless state (Odderon), in addition to the presence of a two-gluon colorless state (Pomeron), could significantly impact the behavior of the $\rho$-parameter at high energies. Consequently, measurements of the parameter $\rho$ at the LHC play a crucial role in understanding these intricate dynamics comprehensively.

After introducing the Odderon, the eikonal scheme demonstrates a slight advantage over the $U$-matrix scheme, mirroring the scenario where the Pomeron is the sole asymptotically dominant entity. Specifically, within the eikonal scheme, predictions for the parameter $\rho^{pp}$ at 13 TeV for Model III with Ensemble A, Model III with Ensemble T, Model IV with Ensemble A, and Model IV with Ensemble T are 0.126, 0.122, 0.127, and 0.128, respectively. In contrast, the corresponding predictions within the $U$-matrix scheme are 0.131, 0.135, 0.131, and 0.139, respectively. 
These predictions exhibit a slight increase compared to those obtained in the analysis without the Odderon, where the behavior of the $pp$ and $\bar{p}p$ channels is equivalent at high energies. However, the introduction of Odderon introduces a distinction between the $pp$ and $\bar{p}p$ channels: the tiny increase in predictions for the $pp$ channel is accompanied by a slightly more noticeable decline for the $\bar{p}p$ channel.
Consequently, for the $\bar{p}p$ channel within the eikonal scheme, the predictions for the parameter $\rho^{\bar{p}p}$ at 13 TeV are as follows: 0.123 for Model III with Ensemble A, 0.120 for Model III with Ensemble T, 0.123 for Model IV with Ensemble A, and 0.126 for Model IV with Ensemble T. The corresponding predictions within the $U$-matrix scheme are 0.125, 0.132, 0.126, and 0.136, respectively. A discernible correlation emerges between the values of $\rho$ and the ratio $\beta^{2}_{\mathbb{O}}(0)/\beta^{2}_{\mathbb{P}}(0)$: generally, a higher value of $\beta^{2}_{\mathbb{O}}(0)/\beta^{2}_{\mathbb{P}}(0)$ tends to correspond to a lower value of $\rho^{\bar{p}p}$. This correlation is a reliable indicator of the impact of the Odderon on interaction dynamics.

Continuing with our analysis incorporating the Odderon, we observe that for an Odderon with a phase factor $\xi_{\Bbb O}= +1$, all eight $\beta_{\mathbb{O}}(0)$ values obtained — four from the eikonal scheme and four from the $U$-matrix scheme — are extremely small (less than $1\times 10^{-8}$) and consistent with zero (errors significantly surpassing central values). Consequently, the remaining parameters assume values very closely resembling the scenario where the Pomeron dominates the scattering amplitude. Hence, one of the key outcomes of our study provides a noteworthy phenomenological indication: the Odderon phase is well-defined and is equal to $\xi_{\Bbb O}= -1$. The significance of the phase $\xi_{\Bbb O}$ becomes apparent when we analyze the contributions of both the Pomeron and the Odderon to the scattering amplitude. The contribution of the Pomeron is primarily imaginary at small $t$ as it dominates the total hadronic cross sections (as indicated by equation (\ref{equation02}), derived from the optical theorem). The Feynman rule for Pomeron exchange is proportional to
\begin{eqnarray}
i\left( \frac{-is}{s_{0}} \right)^{\alpha_{\Bbb P}(t)-1}  ,
\end{eqnarray}
where $s_{0}$ is an energy scale. In turn, the Feynman rule for Odderon exchange is proportional to
\begin{eqnarray}
  \xi_{\Bbb O} \left( \frac{-is}{s_{0}} \right)^{\alpha_{\Bbb O}(t)-1}
  \label{phase003}
\end{eqnarray}
where $\xi_{\Bbb O}$, as previously mentioned, cannot be predetermined. Thus, the amplitude associated with the Odderon is predominantly real. The factor $-i$ on the left of equation (\ref{phase003}) comes from the fact that any pole with $C=-1$ leads to an additional phase $e^{i\pi /2}$ in the amplitude compared to a pole with $C=+1$. We see that $\xi_{\Bbb O}$ can be associated with the Odderon propagator signal.

Despite the unprecedented phenomenological result indicating a well-defined phase factor for the Odderon, it is essential to note that this outcome was derived in the presence of secondary Reggeons. In this scenario, even though Pomeron and Odderon dominate at high energies, we need to rule out the possibility of a subtle interference, particularly at lower energies, between the contributions of Odderon and Reggeons with parity $C=-1$ as the cause of the observed manifestation of the phase factor $\xi_{\Bbb O}= -1$. An ongoing analysis focusing solely on high-energy data, considering exclusively the contributions from Pomeron and Odderon, is imperative to ascertain the stability of the Odderon phase factor.

It is important to note that while introducing an Odderon with a trajectory $\alpha_{\Bbb O}(t) = 1$ does not fully account for the $\rho$ data at $\sqrt{s}=$ 13 TeV. However, it does improve the quality of the overall fit, as evidenced by a lower value of $\chi^{2}$ of the global analysis. A similar outcome was recently reported in reference \cite{lrk2024a}, where the influence of the same type of Odderon was studied by focusing exclusively on differential cross-section data within a two-channel eikonal model. In both studies, only one free parameter related to the Odderon is considered, specifically the Odderon coupling $\beta_{\Bbb O}(0)$. This approach contrasts with that of reference \cite{nicol001}, which proclaims the discovery of the Odderon. In that work the authors focus exclusively on forward quantities ($\sigma_{tot}$ and $\rho$) and utilize a model with three free parameters associated with the so-called Maximal Odderon (MO), providing greater flexibility to the fit. The MO model would require at least six free parameters to describe the differential cross-section data analyzed in this work. Moreover, it is important to note that the MO lacks the structure predicted by  QCD, unlike the Odderon model employed in our study.

Moreover, our analyses have yet to conclusively determine which unitarization scheme might be the most natural at high energies. In a recent study through a fitting procedure to total, elastic, inelastic, and single-diffractive cross-section data (for both $pp$ and $\bar{p}p$ channels), it was discerned that diffractive data exhibit a preference for the $U$-matrix unitarization scheme \cite{cudell01}.
Nevertheless, in two key aspects, our analysis differs from the one presented in reference \cite{cudell01}. Firstly, our choice of experimental data sets diverged as we independently scrutinized the TOTEM and ATLAS data, exploring discrepancies among the various ensembles. This contrasts with reference \cite{cudell01}, where the data are collectively examined. Secondly, a notable distinction lies in our model, which employs a one-channel amplitude, in contrast to the two-channel amplitude approach employed in \cite{cudell01}. The latter approach incorporates the $s$-channel unitarity with elastic and a low mass intermediate state $N^{*}$, accounting for the diffractive proton excitation $p \to N^{*}$.
Integrating diffractive dissociation $p \to N^{*}$ into our model can be achieved through the Good-Walker formalism, a methodology we have already employed in references \cite{broilo001,kmr002,kmr003}. Ongoing investigations involving a two-channel model are underway, focusing on the study of eikonal and $U$-matrix unitarization schemes within the context of our analysis.

\section*{Acknowledgment}

The authors thank M.~G.~Ryskin for reviewing the manuscript and engaging in valuable discussions and V.~Petrov for helpful comments. This research was partially supported by the Agencia Nacional de Investigaci\'on e Innovaci\'on under the project ANII-FCE-166479 and by the Conselho Nacional de Desenvolvimento Cient\'{\i}fico e Tecnol\'ogico under Grant No. 307189/2021-0.


\begin{thebibliography}{99}




  

\bibitem{regge001} V.~Barone and E.~Predazzi, {\it High-Energy Particle Diffraction} (Springer-Verlag, Berlin, 2002).

\bibitem{regge002} S.~Donnachie, G.~Dosch, P.~Landshoff, and O.~Nachtmann, {\it Pomeron Physics and QCD} (Cambridge University Press, Cambridge, United Kingdom, 2002).

\bibitem{regge003} J.~R.~Forshaw and D.~A.~Ross, {\it Quantum Chromodynamics and the Pomeron} (Cambridge University Press, Cambridge, United Kingdom, 1997).
   
\bibitem{cheng001} H.~Cheng and T.~T.~Wu, Phys. Rev. Lett. {\bf 24}, 1456 (1970).

\bibitem{cheng002} C.~Bourrely, J.~Soffer, and T.~T.~Wu, Phys. Rev. Lett. {\bf 54}, 757 (1985).

\bibitem{cheng003} H.~Cheng and T.~T.~Wu, {\it Expanding Protons: Scattering at High Energies} (MIT
Press, Cambridge, MA, 1987).

\bibitem{pdg001}  R.~L.~Workman  {\it et al.} (Particle Data Group), {\it The review of particle physics}, Prog. Theor. Exp. Phys. {\bf 2022}, 083C01 (2022).
  
\bibitem{pirner001} A.~I.~Shoshi, F.~D.~Steffen, and H.~J.~Pirner, Nucl. Phys. A {\bf 709}, 131 (2002).

\bibitem{broilo001} M.~Broilo, D.~A.~Fagundes, E.~G.~S.~Luna, and M.~Pel\'aez, Phys. Rev. D {\bf 103}, 014019 (2021).

\bibitem{landshoff1a} A.~Donnachie and P.~V.~Landshoff, Phys. Lett. B {\bf 798}, 135008 (2019).

\bibitem{landshoff1b} A.~Donnachie and P.~V.~Landshoff, Phys. Lett. B {\bf 831}, 137199 (2022).
  
\bibitem{giffon001} M.~Giffon, E.~Martynov, and E.~Predazzi, Z. Phys. C {\bf 76}, 155 (1997).  
  
\bibitem{cudell01} A.~Vanthieghem, A.~Bhattacharya, R.~Oueslati, and J.~R.~Cudell, JHEP {\bf 09}, 005 (2021).

\bibitem{cudell02} J.~R.~Cudell, E.~Predazzi, and O.~V.~Selyugin, Phys. Rev. D {\bf 79}, 034033 (2009).

\bibitem{selyugin01} O.~V.~Selyugin, J.~R.~Cudell, and E.~Predazzi, Eur. Phys. J. Spec. Top. {\bf 162}, 37 (2008).

\bibitem{anselm001} A.~A.~Anselm and V.~N.~Gribov, Phys. Lett. B {\bf 40}, 487 (1972).

\bibitem{kmr001} V.~A.~Khoze, A.~D.~Martin, and M.~G.~Ryskin, Eur. Phys. J. C {\bf 18}, 167 (2000).

\bibitem{kmr002} E.~G.~S.~Luna, V.~A.~Khoze, A.~D.~Martin, and M.~G.~Ryskin, Eur. Phys. J. C {\bf 59}, 1 (2009).

\bibitem{kmr003} E.~G.~S.~Luna, V.~A.~Khoze, A.~D.~Martin, and M.~G.~Ryskin, Eur. Phys. J. C {\bf 69}, 95 (2010).

\bibitem{kmr004} V.~A.~Khoze, A.~D.~Martin, and M.~G.~Ryskin, Nucl. Phys. B Proc. Suppl. {\bf 99}, 213 (2001).

\bibitem{finkeistein01} J.~Finkelstein, H.~M.~Fried, K.~Kang, and C.~-I.~Tan, Phys. Lett. B {\bf 232}, 257 (1989).
  
\bibitem{martynov01} E.~S.~Martynov, Phys. Lett. B {\bf 232}, 367 (1989).

\bibitem{martynov02} S.~V.~Akkelin and E.~S.~Martynov, Sov. J. Nucl. Phys. {\bf 53}, 1007 (1991).  

\bibitem{martynov03} S.~V.~Akkelin and E.~S.~Martynov, Sov. J. Nucl. Phys. {\bf 55}, 1555 (1992).

\bibitem{low01} F.~E.~Low, Phys. Rev. D {\bf 12}, 163 (1975).

\bibitem{nussinov01} S.~Nussinov, Phys. Rev. Lett. {\bf 34}, 1286 (1975). 

\bibitem{gunion01} J.~F.~Gunion and D.~E.~Soper, Phys. Rev. D {\bf 15}, 2617 (1977).

\bibitem{ryskin01} E.~M.~Levin and M.~G.~Ryskin, Sov. J. Nucl. Phys. {\bf 34}, 619 (1981).

\bibitem{richards01} D.~G.~Richards, Nucl. Phys. B {\bf 258}, 267 (1985).

\bibitem{bopsin01} G.~B.~Bopsin, E.~G.~S.~Luna, A.~A.~Natale, and M.~Pel\'aez, Phys. Rev. D {\bf 107}, 114011 (2023).

\bibitem{ewerz01} C.~Ewerz, arXiv:hep-ph/0306137.

\bibitem{bfkl01} V.~S.~Fadin, E.~A.~Kuraev, and L.~N.~Lipatov, Phys. Lett. B {\bf 60}, 50 (1975).
  
\bibitem{bfkl02} L.~N.~Lipatov, Sov. J. Nucl. Phys. {\bf 23}, 338 (1976).
  
\bibitem{bfkl03} E.~A.~Kuraev, L.~N.~Lipatov, and V.~S.~Fadin, Sov. Phys. JETP {\bf 44}, 443 (1976).

\bibitem{bfkl04} E.~A.~Kuraev, L.~N.~Lipatov, and V.~S.~Fadin, Sov. Phys. JETP {\bf 45}, 199 (1977).

\bibitem{bfkl05} Y.~Y.~Balitsky and L.~N.~Lipatov, Sov. J. Nucl. Phys. {\bf 28}, 822 (1978).

\bibitem{bkp01} J.~Bartels, Nucl. Phys. B {\bf 175}, 365 (1980).

\bibitem{bkp02} T.~Jaroszewicz, Acta Phys. Polon. B {\bf 11}, 965 (1980).
  
\bibitem{bkp03} J.~Kwiecinski and M.~Praszalowicz, Phys. Lett. B {\bf 94}, 413 (1980).

\bibitem{bartels10} J.~Bartels, L.~N.~Lipatov, and G.~P.~Vacca, Phys. Lett. B {\bf 477}, 178 (2000).

\bibitem{antchev001} G.~Antchev {\it et al.}, Europhys. Lett. {\bf 101}, 21002 (2013).

\bibitem{antchev002} G.~Antchev {\it et al.}, Nucl. Phys. B {\bf 899}, 527 (2015).

\bibitem{TOTEM001} G.~Antchev {\it et al.}, Europhys. Lett. {\bf 95}, 41001 (2011).

\bibitem{TOTEM005} G.~Antchev {\it et al.}, Eur. Phys. J. C {\bf 76}, 661 (2016).

\bibitem{TOTEM008} G.~Antchev {\it et al.}, Eur. Phys. J. C {\bf 79}, 785 (2019).  

\bibitem{TOTEM010} G.~Antchev {\it et al.}, Eur. Phys. J. C {\bf 79}, 861 (2019).

\bibitem{atlas001} G.~Aad {\it et al.}, Nucl. Phys. B {\bf 889}, 486 (2014).

\bibitem{atlas002} M.~Aaboud {\it et al.}, Phys. Lett. B {\bf 761}, 158 (2016). 

\bibitem{ATLAS01} G.~Aad {\it et al.}, Eur. Phys. J. C {\bf 83}, 441 (2023).

\bibitem{petrov01} V.~Petrov and N.~P.~Tkachenko, Phys. Part. Nucl. {\bf 54}, 1152 (2023).

\bibitem{petrov02} V.~Petrov and N.~P.~Tkachenko, Nucl. Phys. A {\bf 1042}, 122807 (2024).  

\bibitem{luna011} E.~G.~S.~Luna and M.~J.~Menon, arXiv:hep-ph/0105076.

\bibitem{luna012a} E.~G.~S.~Luna and M.~J.~Menon, Phys. Lett. B {\bf 565}, 123 (2003).
  
\bibitem{luna012b} E.~G.~S.~Luna, M.~J.~Menon, and J.~Montanha, Nucl. Phys. A {\bf 745}, 104 (2004).

\bibitem{luna012c} E.~G.~S.~Luna, M.~J.~Menon, and J.~Montanha, Braz. J. Phys. {\bf 34}, 268 (2004).  
  
\bibitem{blockcahn} M.~M.~Block and R.~N.~Cahn, Rev. Mod. Phys. {\bf 57}, 563 (1985).

\bibitem{goulianos001} R.~J.~M.~Covolan, J.~Montanha, and K.~Goulianos, Phys. Lett. B {\bf 389}, 176 (1996).  

\bibitem{kmrepj01} V.~A.~Khoze, A.~D.~Martin, and M.~G.~Ryskin, Eur. Phys. J. C {\bf 73}, 2503 (2013).

\bibitem{kmrepj02} V.~A.~Khoze, A.~D.~Martin, and M.~G.~Ryskin, Phys. Lett. B {\bf 784}, 192 (2018).  

\bibitem{lev01} E.~Gotsman, E.~Levin, and U.~Maor, Int. J. Mod. Phys. A {\bf 30}, 1542005 (2015).

\bibitem{lev02} E.~Gotsman, E.~Levin, and U.~Maor, Phys. Rev. D {\bf 87}, 071501(R) (2013).

\bibitem{gribov01} V.~N.~Gribov, Sov. Phys. JETP {\bf 15}, 873 (1962).

\bibitem{gribov02} V.~N.~Gribov, Sov. Phys. JETP {\bf 26}, 414 (1968).

\bibitem{gribov03} V.~N.~Gribov, Sov. J. Nucl. Phys. {\bf 9}, 369 (1969).

\bibitem{baker01} M.~Baker and K.~A.~Ter-Martirosyan, Phys. Rep. {\bf 28}, 1 (1976).

\bibitem{auger} P.~Abreu {\it et al.}, Phys. Rev. Lett. {\bf 109}, 062002 (2012).

\bibitem{TA} R.~U.~Abbasi {\it et al.}, Phys. Rev. D {\bf 92}, 032007 (2015).

\bibitem {delta1} S.~M.~Roy and V.~Singh, Phys. Lett. B {\bf 32}, 50 (1970).

\bibitem {delta2} R.~J.~Eden, Rev. Mod. Phys. {\bf 43}, 15 (1971).

\bibitem {pmrchuk1} R.~J.~Eden, Phys. Rev. Lett. {\bf 16}, 39 (1966).

\bibitem {pmrchuk2} G.~Grunberg and T.~N.~Truong,  Phys. Rev. Lett. {\bf 31}, 63 (1973).

\bibitem {pmrchuk3} G.~Grunberg and T.~N.~Truong, Phys. Rev. D {\bf 9}, 2874 (1974).

\bibitem{lunphys01} M.~Broilo, D.~A.~Fagundes, E.~G.~S.~Luna, and M.~J.~Menon, Phys. Lett. B {\bf 799}, 135047 (2019).

\bibitem{lunphys02} M.~Broilo, E.~G.~S.~Luna, and M.~J.~Menon, Phys. Rev. D {\bf 98}, 074006 (2018).

\bibitem{lrk2024a} E.~G.~S.~Luna, M.~G.~Ryskin, and V.~A.~Khoze, Phys. Rev. D {\bf 110}, 014002 (2024).

\bibitem{nicol001} E.~Martynov and B.~Nicolescu, Phys. Lett. B {\bf 778}, 414 (2018).


  

\end{thebibliography}
\end{document}